\documentclass[12pt]{iopart}

\usepackage{graphicx}
\usepackage{wasysym} %for diameter symbol
\usepackage{hyperref}
\usepackage{color}
\begin{document}

\def\mydm{\Delta m^2_{41}}
\def\mysin{\sin^2 2\theta_{ee}}
\def\oscillationspars{$\mydm$, $\mysin$}
\def\antiparticle{\tilde}
\def\antinu{$\tilde{\nu_e}$}
\def\clsmethod{CL$_s$}
\def\cl{\mathrm{CL}}
\def\todo{\textcolor{blue}}

\definecolor{goodcolor}{rgb}{.9 ,.01 ,.01}%
\definecolor{badcolor}{rgb}{.01 ,.2 ,.9}%
\definecolor{normalcolor}{rgb}{.0 ,.0 ,.0}%

\title[Review of sterile neutrino searches at very short-baseline reactor experiments]{Review of sterile neutrino searches at very short-baseline reactor experiments}

\author{Mikhail Danilov}

\address{Lebedev Physical Institute of the Russian Academy of Sciences,\\
  53 Leninskiy Prospekt, Moscow, Russia}
\ead{danilov@lebedev.ru}
\vspace{10pt}
%\begin{indented}
%\item[]August 2017
%\end{indented}

\begin{abstract}
Search for New Physics beyond the Standard Model is the main direction in particle physics nowadays. There are several experimental hints of New Physics. The most statistically significant (5-6$\sigma$) are the hints of eV mass scale sterile neutrinos. They come from $\antiparticle\nu_e$ disappearance in reactor experiments, $\nu_e$ disappearance in experiments with very powerful radioactive sources, and electron (anti)neutrino appearance in the muon (anti)neutrino beams. Very important results in this field were obtained in 2021 by the BEST, MicroBooNE, and Neutrino-4 collaborations as well as by several other experiments. However, the situation is still or maybe even more controversial.   We review these indications of New Physics and prospects for the next few years with the emphasis on reactor experiments.   

\end{abstract}

%
% Uncomment for keywords
\vspace{2pc}
\noindent{\it Keywords}: neutrino, reactor, sterile neutrino, neutrino oscillations, Beyond Standard Model

% Uncomment for Submitted to journal title message
%\submitto{\JPA}
%
% Uncomment if a separate title page is required
%\maketitle
% 
% For two-column output uncomment the next line and choose [10pt] rather than [12pt] in the \documentclass declaration
%\ioptwocol
%

\section{Introduction}

The Standard Model is probably the most elaborate theory of Matter ever developed (for a recent review see e.g.~\cite{PDG}). It describes practically everything that we observe very often with very high precision. The final confirmation of the SM was the discovery of the famous Higgs Boson at the Large Hadron Collider (LHC) in 2012~\cite{CMS, ATLAS}. Still, it is commonly believed that SM is not  the ultimate theory. First of all, it does not describe what we do not see, namely, the Dark Matter that is about 6 times more abundant in the Universe than the ordinary matter that is described by SM~\cite{DM-Planck2018}. The dominance of matter over antimatter in the Universe is also not explained.  Fine tuning of the model’s parameters is required to explain a relatively small mass of the Higgs Boson. And the number of these free model parameters is very large. SM does not predict the masses of fundamental particles, the couplings between quarks of different generations, and the strength of the weak, electromagnetic and strong interactions. It does not explain  why there are just 3 generations of quark and leptons. Finally, the SM does not include gravity.
Therefore, search for phenomena beyond SM is now the main direction in particle physics. 

The are several indications of Physics beyond SM. The long-standing discrepancy between the very precise experimental measurements (relative accuracy is 0.35~ppm) and theoretical predictions (relative accuracy is 0.37~ppm) for the anomalous muon magnetic moment has about 4.2 standard deviation ($\sigma$) statistical significance~\cite{PhysRevLett.126.141801}. A considerable improvement in the experimental accuracy is expected in the near future.

There are several discrepancies at the level of 3$\sigma$ between experimental data and SM predictions in beauty hadron decays.
%~\cite{PDG}. 
Probably the most remarkable is the difference in the probabilities of decays with $\mu^{+}\mu^{-}$ and $e^+e^-$ in the final state which should be equal in SM (see~\cite{arXiv:2110.09501v2} and references therein). 
Again, a better accuracy is required in order to make definite conclusions. A considerable improvement in experimental accuracy is expected in this field in experiments at LHC as well as at the SuperB Factory at KEK.
%~\cite{KEK-B}. 

Finally, several indications of New Physics beyond SM exist in the neutrino sector. Discovery of neutrino oscillations~\cite{SuperK,PhysRevLett.87.071301} demonstrated that neutrinos have masses. This fact can be considered as a sign of New Physics since neutrino are massless in SM. However, it is possible to extend SM and include neutrino masses to it. There is another hint of New Physics in the neutrino sector – sterile neutrinos. There are several experimental indications of existence of sterile neutrinos with masses of the order of 1~eV.
A detailed discussion of these indications and future prospects for sterile neutrino searches with the emphasis on reactor experiments are the main topics of the present paper. 

\section{Hints for sterile neutrinos}
Neutrino oscillations were discovered by the Super Kamikoande experiment~\cite{SuperK} in case of atmospheric neutrinos and several solar neutrino experiments~\cite{Davis:2003xx,SAGE,GALEX} with the final prove by SNO~\cite{PhysRevLett.87.071301}.
The observation of neutrino oscillations demonstrates that neutrino have masses and that the lepton mixing matrix (Pontecorvo, Maki, Nakagava, Sakata matrix) \cite{Pontecorvo226, MNS227, Pontecorvo228} is non trivial. 
This matrix connects the neutrino flavor eigenstates  $\nu_e$, $\nu_{\mu}$, and $\nu_{\tau}$ with the neutrino  mass eigenstates $\nu_1$, $\nu_2$, and $\nu_3$.
Since the measured neutrino mass-squared differences demonstrate a hierarchical structure %($\Delta^2_{atm}=?\pm?eV^2,\Delta^2_{solar}=?\pm?eV^2$~\cite{PDG})
($|\Delta m^2_{atm}| \approx 2.5 \times 10^{-3}$~eV$^2$,
$\Delta m^2_{solar}=7.53\pm0.18 \times 10^{-5}$~eV$^2$~\cite{PDG})
 the probabilities of  neutrino flavor  changes from $\alpha$ to $\beta$ are often well approximated by a two-neutrino formula(see e.g.~\cite{arXiv:2111.07586})
 
\begin{equation}
\label{eq:osc_ab}
Prob(\nu_\alpha \rightarrow \nu_\beta) \approx 
\sin^22\theta_{\alpha\beta}\sin^2\left(\frac {1.27\Delta m^2_{ij} [\mathrm{eV}^2] L[\mathrm m]}{E_\nu [\mathrm{MeV}]}\right),
\end{equation}
with only one mixing angle $\theta_{\alpha\beta}$ and one difference of neutrino masses squared $ \Delta m^2_{ij}=m_i^2-m_j^2$ of the relevant neutrino mass states.
%, where
In this expression $E_{\nu}$ and L are the neutrino energy and the path length. The survival probability is than given by
\begin{equation}
\label{eq:1-osc_ee}
Prob(\nu_\alpha \rightarrow \nu_\alpha) \approx 1 - 
\sin^22\theta_{\alpha\alpha}\sin^2\left(\frac {1.27\Delta m^2_{ij} [\mathrm{eV}^2] L[\mathrm m]}{E_\nu [\mathrm{MeV}]}\right).
\end{equation}
At large distances the oscillation pattern is smeared out because of finite accuracy of energy and/or distance measurements and only the initial neutrino flux is reduced:
\begin{equation}
\label{eq:1-osc_ee_L}
Prob(\nu_\alpha \rightarrow \nu_\alpha) \approx 1 - 0.5
\sin^22\theta_{\alpha\alpha}.
\end{equation}
Oscillations of the three known neutrino flavors have been measured with a good precision. The two mass-squared differences and three angles describing such oscillations are well known by now\cite{PDG}. %Additional light active neutrinos are excluded by the measurements of the Z boson decay width \cite{PDG}. Nevertheless, existence of additional sterile neutrinos is not excluded. Moreover, several effects observed with about $3\sigma$ significance level can be explained by active-sterile neutrino oscillations.
Z boson  can decay into a neutrino-antineutrino pair. Therefore
measurements of the Z boson decay width can be used to determine the  number of light active neutrinos to be equal to 3~\cite{2006257}. However, additional sterile neutrinos are allowed. They are singlets of the SM gauge group and do not interact directly with gauge bosons.
Usually only one sterile neutrino $\nu_s$ is considered which consists mainly of the heavy $\nu_4$ neutrino mass state while the 3 active neutrinos $\nu_e$, $\nu_{\mu}$, and $\nu_{\tau}$ are mainly composed of light neutrino  mass states $\nu_1$, $\nu_2$, and $\nu_3$. This is the so called 3+1 sterile-active neutrino mixing  model which we will call the 4$\nu$ model for brevity.
 In this model  $\theta_{ee}\approx\theta_{14}$ since all relevant angles are small.
Addition of more sterile neutrinos usually does not change results considerably in comparison with the 3+1 model.

Sterile neutrinos appear naturally in many extensions of SM. Moreover, there are several experimental hints of their existence.
The GALEX and SAGE solar neutrino Gallium experiments performed calibrations with very powerful radioactive sources.
%and reported the ratio of numbers of 
They observed only $(88\pm 5)\%$ $\nu_e$ events of the expected number~\cite{SAGE}.
The deficit of $\nu_e$ events in the calibration runs of the SAGE and GALEX experiments~\cite{SAGE, GALEX} (``Galium Anomaly''(GA)~\cite{Ga}) can be explained by electron neutrino to sterile neutrino oscillations at very short distances with quite a large $\sin^22\theta_{ee}\approx0.24$ according to equation~(\ref{eq:1-osc_ee_L}).

Analogously the 6\% deficit of  $\antiparticle\nu_e$ in comparison with the re-evaluated reactor fluxes~\cite{Huber,Mueller} (``Reactor Antineutrino Anomaly''(RAA)~\cite{Mention2011}) can be explained by the active-sterile neutrino oscillations again at very short distances. The required short distances of the oscillations imply that the mass-squared difference $ \Delta m^2$ should be about 1~eV$^2$, much larger than the two known mass-squared differences. Therefore an additional neutrino is needed and,  as explained above, it should be sterile.
Since the mass-squared difference  between a new and known neutrinos is much larger than
the mass-squared differences between known neutrinos the  oscillations  between known and sterile neutrinos can be well described by a two-neutrino oscillation formula~(\ref{eq:osc_ab}) with only one $\Delta m^2_{41}$.
 RAA leads to the  best-fit values for sterile neutrino parameters of $\Delta m^2_{41} = 2.3$ eV$^2$ and $\sin^2 2\theta_{ee}$ = 0.14.

More recent estimates of the $\antiparticle\nu_e$ fluxes from reactors~\cite{PhysRevC.100.054323, PhysRevLett.123.022502} have not solved the problem with RAA  since one of them~\cite{PhysRevLett.123.022502} predicts a smaller $\antiparticle\nu_e$ flux while the other one~\cite{PhysRevC.100.054323} predicts a larger $\antiparticle\nu_e$ flux in comparison with the Huber-Mueller (H-M) model~\cite{Huber,Mueller}.
The Daya Bay~\cite{An:2017osx} and RENO~\cite{PhysRevLett.122.232501} collaborations have measured the $^{235}$U contribution to the reactor $\antiparticle\nu_e$ flux multiplied by the Inverse Beta Decay (IBD) reaction  cross section. The IBD process is used to detect $\antiparticle\nu_e$ at reactors.
The results are about 10\% smaller than the H-M model predictions. 

The H-M model converts the measurements of the  beta spectra from thermal neutron fission products of $^{235}$U, $^{239}$Pu, and $^{241}$Pu at ILL ~\cite{SCHRECKENBACH1981251, SCHRECKENBACH1985325, VONFEILITZSCH1982162, HAHN1989365} to the $\antiparticle\nu_e$ spectra.
Recent measurements at the Kurchatov Institute~\cite{Kopeikin:2021rnb,Kopeikin:2021ugh} of the ratio of beta spectra of fission products of $^{235}$U and $^{239}$Pu give $(5.4\pm0.2)\% $ smaller ratio than the ILL results.
%~\cite{SCHRECKENBACH1981251, SCHRECKENBACH1985325,
 %{VONFEILITZSCH1982162, HAHN1989365}. 
 %used for predictions of the $\antiparticle\nu_e$ fluxes from reactors~\cite{Huber,Mueller}.

Smaller values for $^{235}$U contribution to the IBD rate obtained by Daya Bay and  RENO, and the Kurchatov Institute experiment result reduce considerably the significance of RAA.
 There is also a difference between  measured and predicted $\antiparticle\nu_e$ energy spectra. Their ratio has a bump at about 6~MeV\cite{doi:10.1063/1.4915563, doi:10.1007/JHEP10(2014)086, PhysRevLett.116.061801}.
 However, all modern searches for sterile neutrinos measure the ratio of the $\antiparticle\nu_e$ energy spectra at different distances from reactors and therefore they do not rely on the absolute $\antiparticle\nu_e$ flux predictions as well as on the shape of the predicted $\antiparticle\nu_e$ energy spectrum.

Very recently the BEST experiment has confirmed GA~\cite{BEST-2021-arxiv,BEST-2022-arxiv}. The observed deficit of $\nu_e$ events is even larger ($(79\pm 5)\%$ and $(77\pm 5)\%$ in the inner and outer volumes of the detector) and has a significance of above 5$\sigma$~\cite{arXiv:2109.14654v1}. For $\Delta m^2_{41} < 5$eV$^2$ the very large values of $\sin^2 2\theta_{ee}\approx 0.4$ preferred by BEST (see equation (\ref{eq:1-osc_ee_L})) have been already excluded by DANSS~\cite{Alekseev:2018efk,Danilov:2020ucs} and NEOS~\cite{Ko:2016owz}. This will be discussed in the next section. For large $\Delta m^2_{41} > 5$eV$^2$ and $\sin^2 2\theta_{ee}$ the BEST results are in tension with the limits obtained by Daya Bay, Bugey-3 and RENO (see e.g.~\cite{MINOS:2020iqj}) using predictions for the absolute $\antiparticle\nu_e$ flux from reactors including their large uncertainties. A part of the BEST preferred region at large $\Delta m^2$ was already excluded by the PROSPECT~\cite{PhysRevD.103.032001} and STEREO~\cite{PhysRevD.102.052002} experiments.

\begin{figure}[th]%1
%\vspace{-4.2cm}
\centering
\includegraphics[width=0.44\linewidth]{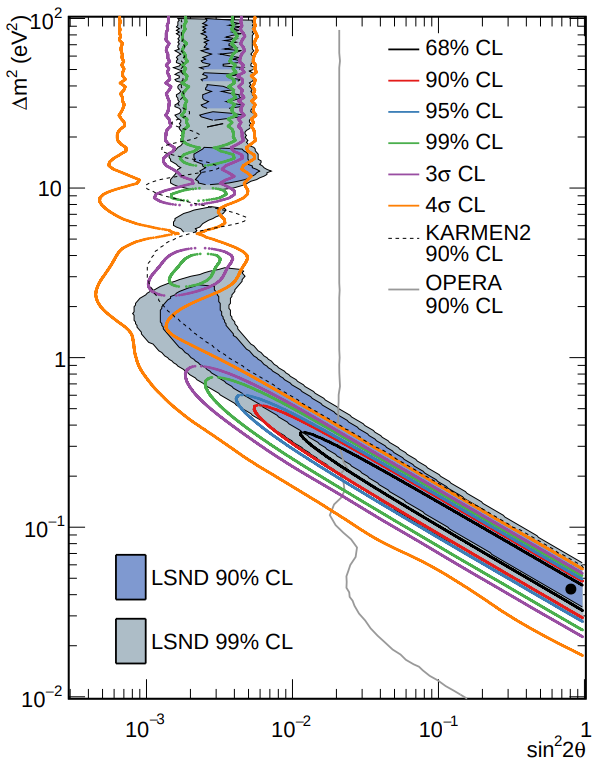}
%\begin{minipage}[b]{15pc}
 \caption{MiniBooNE allowed regions for combined neutrino and antineutrino mode  data sets for events with $200 < E_{\nu} < 3000$~MeV within a two-neutrino oscillation model. The shaded areas show the 90\% and 99\% Confidence Level (C.L.) LSND $\antiparticle{\nu_{\mu}} \to \antiparticle{\nu_e}$ allowed regions. The black point shows the MiniBooNE best-fit point. Also shown are 90\% C.L. limits from the KARMEN~\cite{PhysRevD.65.112001} and OPERA~\cite{Agafonova2018} experiments. Figure is adopted from~\cite{PhysRevD.103.052002}.
}
\label{fig:MiniBooNe}
%\end{minipage}
\end{figure}

 The LSND collaboration obtained 
 evidence for electron antineutrino appearance in the muon antineutrino beams at distances where known neutrino oscillations can not contribute~\cite{LSND_2018}. This signal can be explained by the oscillations of muon antineutrinos to sterile neutrinos followed by oscillation of sterile neutrinos to electron antineutrinos. The corresponding
  mass-squared difference should be larger than $\sim 0.2~$eV$^2$~\cite{LSND_2018} that is much bigger than the mass-squared differences of known neutrino oscillations. The first MiniBooNE checks  of the LSND signal were not conclusive
  ~\cite{PhysRevLett.98.231801, PhysRevLett.102.101802, PhysRevLett.110.161801}.
  Later the MiniBooNE collaboration obtained the $4.8\sigma$ evidence for electron (anti)neutrino appearance in the muon (anti)neutrino beams~\cite{MiniBooNE2_2018, PhysRevD.103.052002}.
The significance of the signal grows to 6.1$\sigma$ if the LSND and MiniBooNE results are combined~\cite{PhysRevD.103.052002}.
The best-fit point in the sterile neutrino parameter space is close to the maximal mixing ($\sin^2 2\theta_{e\mu}=0.807$) and a small mass-square difference of $\Delta m_{14}^2=0.043\mathrm{eV}^2$ (see Figure~\ref{fig:MiniBooNe}). However, this area  is excluded by OPERA~\cite{Agafonova2018} and only a region with larger mass-squared differences  and smaller mixing is still allowed.

The sterile neutrino explanation of the LSND and MiniBooNE results requires mixing of sterile neutrinos both with electron and muon neutrinos. Therefore 
the probability of the electron neutrino appearance in the muon neutrino beam is proportional to the probabilities of the $\nu_e$ disappearance and $\nu_{\mu}$  disappearance ($\sin^22\theta_{e\mu}\approx 0.25 \sin^22\theta_{ee}\sin^22\theta_{\mu\mu}$ for small angles). Hence the strong upper limits on  $\nu_{\e}$ and $\nu_{\mu}$ disappearance lead to strong limits on the $\nu_e$ appearance in $\nu_{\mu}$ beams that contradict the LSND and MiniBooNE results~\cite{Giunti_2018, Dentler_2018}. However one should mention that a weak indication of $\nu_{\mu}$  disappearance was obtained by the Ice Cube experiment ($\sin^2 2\theta_{\mu\mu}=0.10,\Delta m_{14}^2=4.5\mathrm{eV}^2$)~\cite{PhysRevLett.125.141801}.
With such a large mixing parameter appearance and disappearance results are marginally compatible at large $\Delta m_{14}^2$. On the other hand such large mixing is in contradiction with strong limits on $\nu_{\mu}$ disappearance mainly form the MINOS/MINOS+ experiments~\cite{PhysRevLett.122.091803}.

\begin{figure}[th]%1
%\vspace{-4.2cm}
\centering
\includegraphics[width=0.53\linewidth]{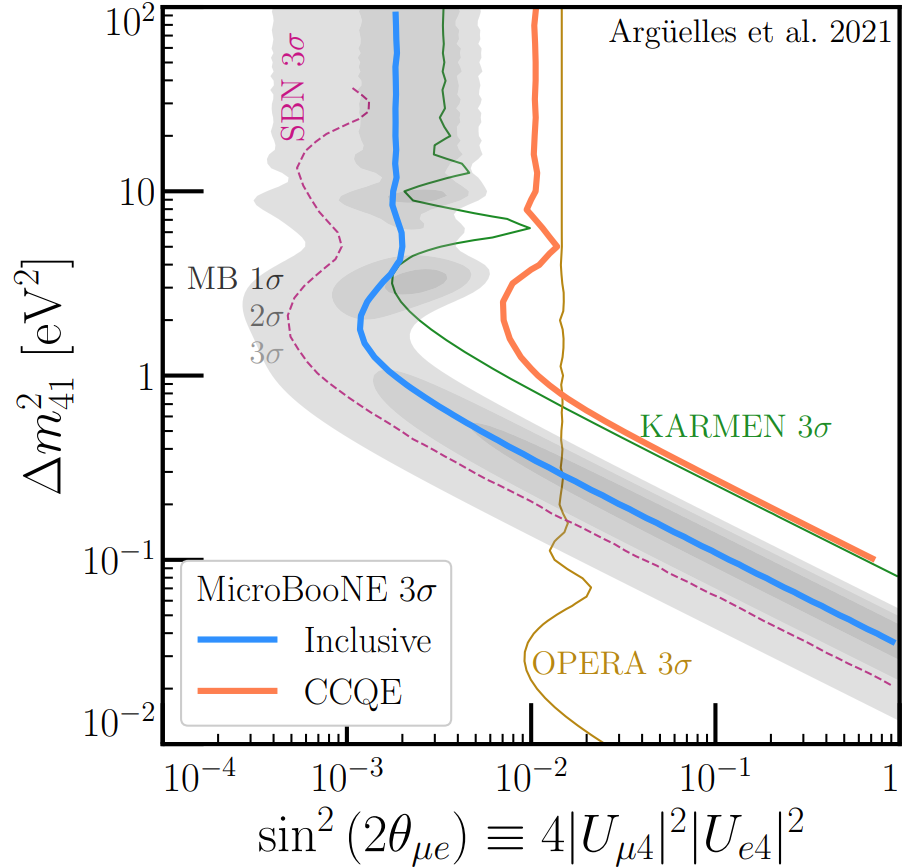}
%\begin{minipage}[b]{15pc}
 \caption{MicroBooNE constraints on the sterile neutrino parameter space at 3$\sigma$ C.L. (blue, inclusive and orange, Charged-Current 
Quasielastic). Also are shown the MiniBooNE 1-, 2-, and 3-$\sigma$ preferred regions in shades of grey~\cite{PhysRevD.103.052002}, the future sensitivity of the three SBN detectors (pink~\cite{doi:10.1146/annurev-nucl-101917-020949}), and existing constraints from KARMEN (green~\cite{PhysRevD.65.112001}) and OPERA (gold~\cite{Agafonova2018}). Figure is adopted from~\cite{2111.10359v1}.
}
\label{fig:MicroBooNe}
%\end{minipage}
\end{figure}

The overall situation with the $\nu_e$ appearance is not yet clear and more results are needed to clarify it.
The MiniBooNE results are scrutinized by the MicroBooNE experiment at the same neutrino beam. MicroBooNE  observed even smaller number of $\nu_{\e}$ events than expected and established upper limits on the possible excess~\cite{MicroBooNE:2021rmx}. However, the sterile neutrino explanation of the MiniBooNE excess is not completely ruled out~\cite{2111.10359v1} (see Figure~\ref{fig:MicroBooNe}).

\begin{figure}[th]%1
%\vspace{-4.2cm}
\centering
\includegraphics[width=0.9\linewidth]{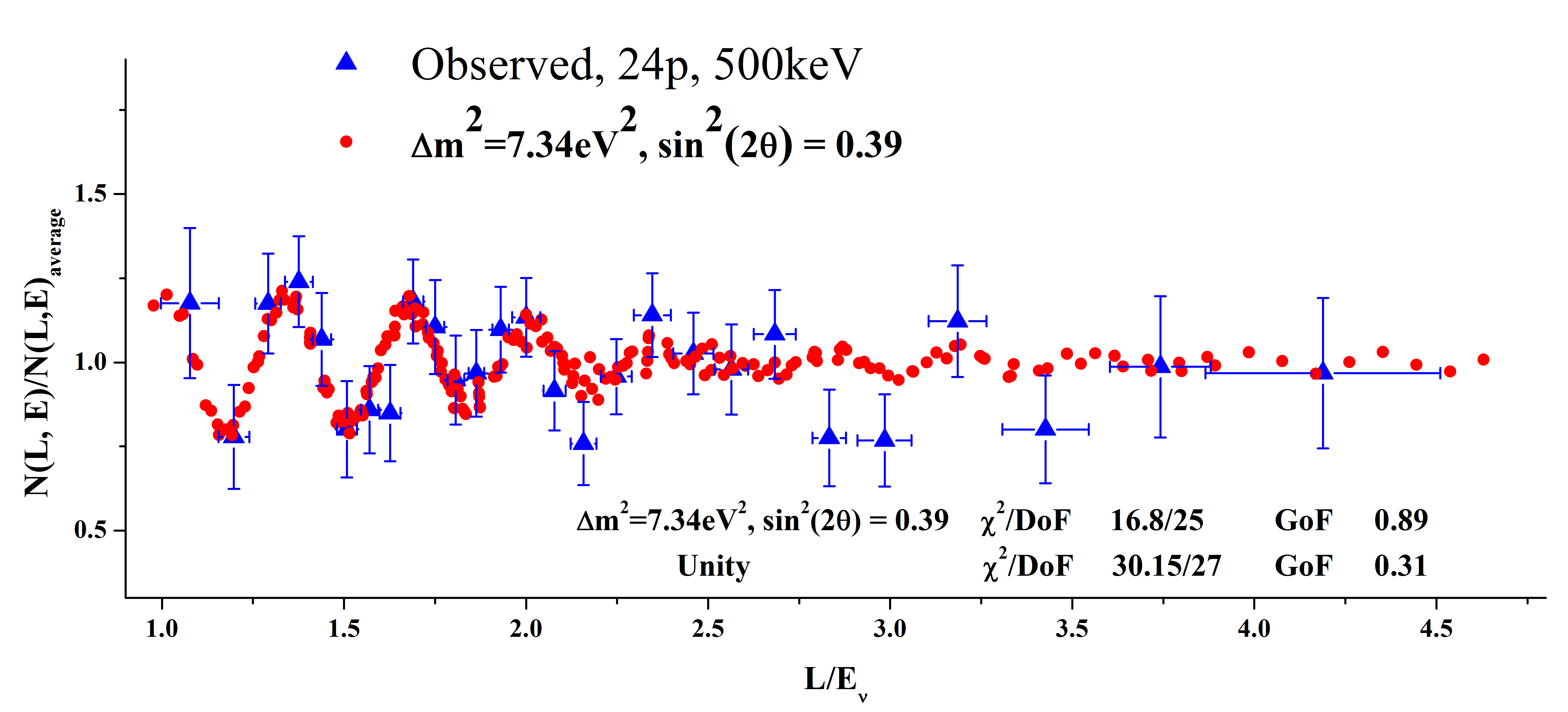}
 \caption{\footnotesize 
Normalized L/E dependence of the IBD rate in the Neutrino-4 experiment (triangles) and theoretical predictions (dots). Figure is adopted from~\cite{Serebrov:2018vdw}.}
 \label{fig:Nu4LE}
\end{figure}

The Neutrino-4 collaboration claimed in 2018 an observation of $\antiparticle\nu_e$ oscillations to sterile neutrinos with very large values of $\Delta m^2_{41} \simeq 7$ eV$^2$ and $\sin^2 2\theta_{ee} \simeq 0.4$
although the significance of the result was only 2.8$\sigma$\cite{Serebrov:2018vdw,Serebrov:2020rhy} (see Figure~\ref{fig:Nu4LE}).

The Neutrino-4 analysis was criticized in~\cite{Danilov:2018dme,Danilov:2020rax,Almazan:2020drb,Giunti:2021iti}. Neutrino-4 initially refused these critical comments~\cite{Serebrov:2020wny,Serebrov:2020yvp} but later took into account two of them~\cite{Neutrino-4-2021}. This reduced the significance of the result by $\approx 0.5\sigma$. The best-fit point for the improved analysis and increased data sample is  $\Delta m^2_{41} = 7.3\pm 1.17$ eV$^2$, $\sin^2 2\theta_{ee} = 0.36\pm0.12_{\rm stat}$~\cite{Neutrino-4-2021}. The statistical significance of the result is 2.7$\sigma$.
The Neutrino-4 claim is in tension with the PROSPECT results~\cite{PhysRevD.103.032001} as well as with the measurements of the absolute $\antiparticle\nu_e$ flux from reactors~\cite{MINOS:2020iqj} and Solar neutrino measurements~\cite{GIUNTI2021136214}. On the other hand the Neutrino-4 results are in a perfect agreement with the recent BEST result~\cite{BEST-2021-arxiv,BEST-2022-arxiv}.

\section{Searches for sterile neutrinos at very short base-line  reactor experiments}
This section is a cardinal update of the presentation at ICCPA-2018~\cite{Danilov:2018dme}.
\subsection{The DANSS Experiment}
The DANSS detector~\cite{DANSS} consists of 2.5 thousand scintillator strips ($1 \times 4 \times 100$~cm${}^3$) with a thin ($\sim 0.2$~mm) Gd-containing reflective surface coating. It is  surrounded with a composite shielding to suppress backgrounds. DANSS is placed on a movable platform under the core of the 3.1~GW$_{\rm th}$  industrial power reactor at the Kalinin Nuclear Power Plant (KNPP) in Russia. The detector distance to the reactor core center is changed from 10.9~m to 11.9~m, and  12.9~m 2-3 times a week. Reactor materials provide  $\sim$~50~m of water equivalent (m.w.e.) shielding that removes the hadron component of the cosmic background and reduces the cosmic muon flux by a factor of 6. The very good suppression of the cosmic background and the high granularity of the detector allow DANSS to achieve a very high signal/background (S/B) ratio of more than 50 (at 10.9~m from the reactor, after model independent subtraction  of the accidental background).
The size of the reactor core is quite big (3.7~m in height and 3.2~m in diameter) which leads to smearing of the oscillation pattern. This drawback is compensated by a high $\antiparticle{\nu_e}$  flux which allows DANSS to detect more than 5 thousand $\antiparticle\nu_e$ at a distance of 10.9~m. 
The energy resolution of the DANSS detector is very modest ($\sigma_E/E \sim 34\%$ at $E=1$~MeV). This leads to additional smearing of the oscillation pattern, comparable with the smearing due to the large reactor core size. 

The IBD reaction
$\antiparticle{\nu}_e + p \rightarrow e^+ + n$ 
is used to detect $\antiparticle{\nu}_e$.
Positrons immediately deposit their energy in the detector and annihilate with production of 2 or 3 gammas with the total energy of 1.02~MeV. Neutrons are slowed down to thermal energies where the capture cross-section is high and then captured by Gd which in turn emits gammas with the total energy of about 8 MeV. Thus the IBD reaction produces two signals, prompt and delayed. The delayed coincidence of these two signals allows to suppress the background drastically. The ${\antiparticle\nu_e}$ energy can be inferred from the positron energy
 $ E_{\antiparticle\nu} \approx E_{e^+} + 1.8~\mathrm{MeV}$ or from the prompt signal energy that includes 1.02~MeV from the annihilation gammas. 

The DANSS experiment compares the positron energy spectra measured with the same detector at the three distances from the reactor core using only relative IBD counting rates. This is the most conservative approach which does not depend on the predicted $\antiparticle\nu_e$ flux and spectrum as well as on the detector efficiency (only short term variations of the efficiency can influence the results).

\begin{figure}[h]
\centering
\includegraphics[width=0.65\textwidth]{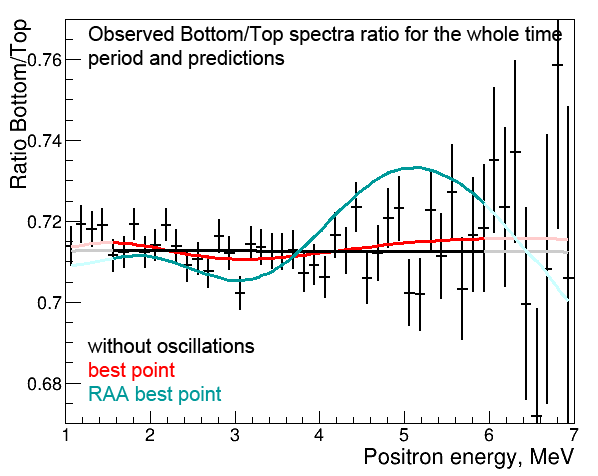}
%\begin{minipage}[b]{16pc}
\caption{ 
Ratio of positron energy spectra measured at the bottom and top detector positions  (statistical errors only). 
 The black curve is the prediction for the 3$\nu$ case, the red curve corresponds to the best fit in the $4\nu$ mixing scenario ($\mysin = 0.014$, $\mydm = 1.3~\rm{eV}^2$) for the full analysis %with 2 separate data taking strategies
, the cyan curve is the expectation for the best-fit point for RAA \cite{Mention2011} ($\mysin=0.14$, $\mydm = 2.3~\rm{eV}^2$).
}
%\end{minipage}
\label{Ratio}
\end{figure}

Figure~\ref{Ratio} 
%\cite{Your talk in Greece?}  
shows the ratio of positron energy spectra at the bottom and top detector positions. It is consistent with the 3$\nu$ hypothesis (black curve). There is no statistically significant evidence for sterile neutrinos. The best 4$\nu$ hypothesis fit (red curve) is only slightly better (1.3$\sigma$).
Hence there is no statistically significant evidence for sterile neutrinos. The best-fit point of RAA  (cyan curve) is clearly excluded at more than 5$\sigma$.  

\begin{figure}[h]%2
\centering
%\begin{minipage}[t]{0.47\linewidth}
\includegraphics[width=0.47\linewidth]{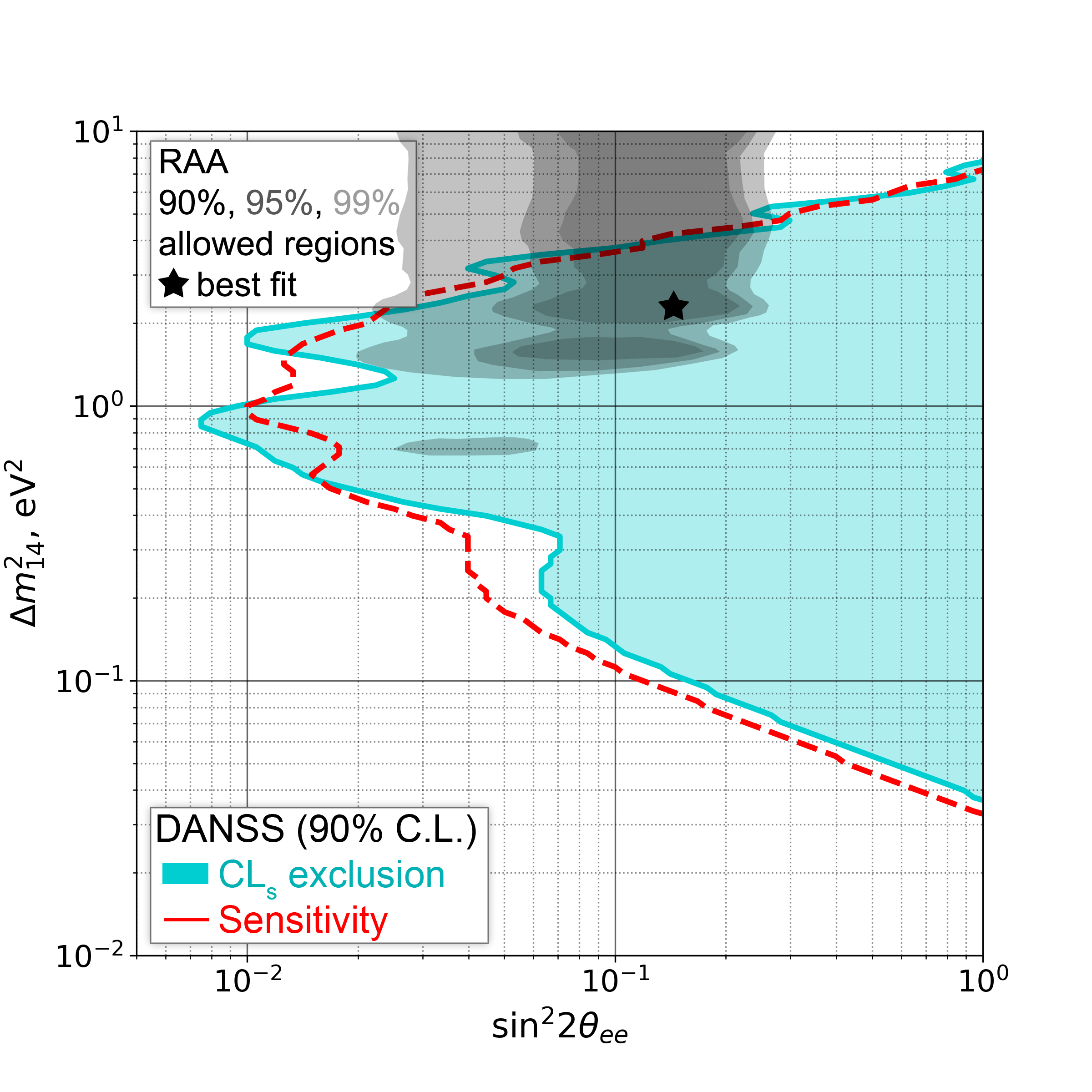}
% \end{minipage}\hspace{2pc}%
%\begin{minipage}[t]{0.47\linewidth}
\includegraphics[width=0.47\linewidth]{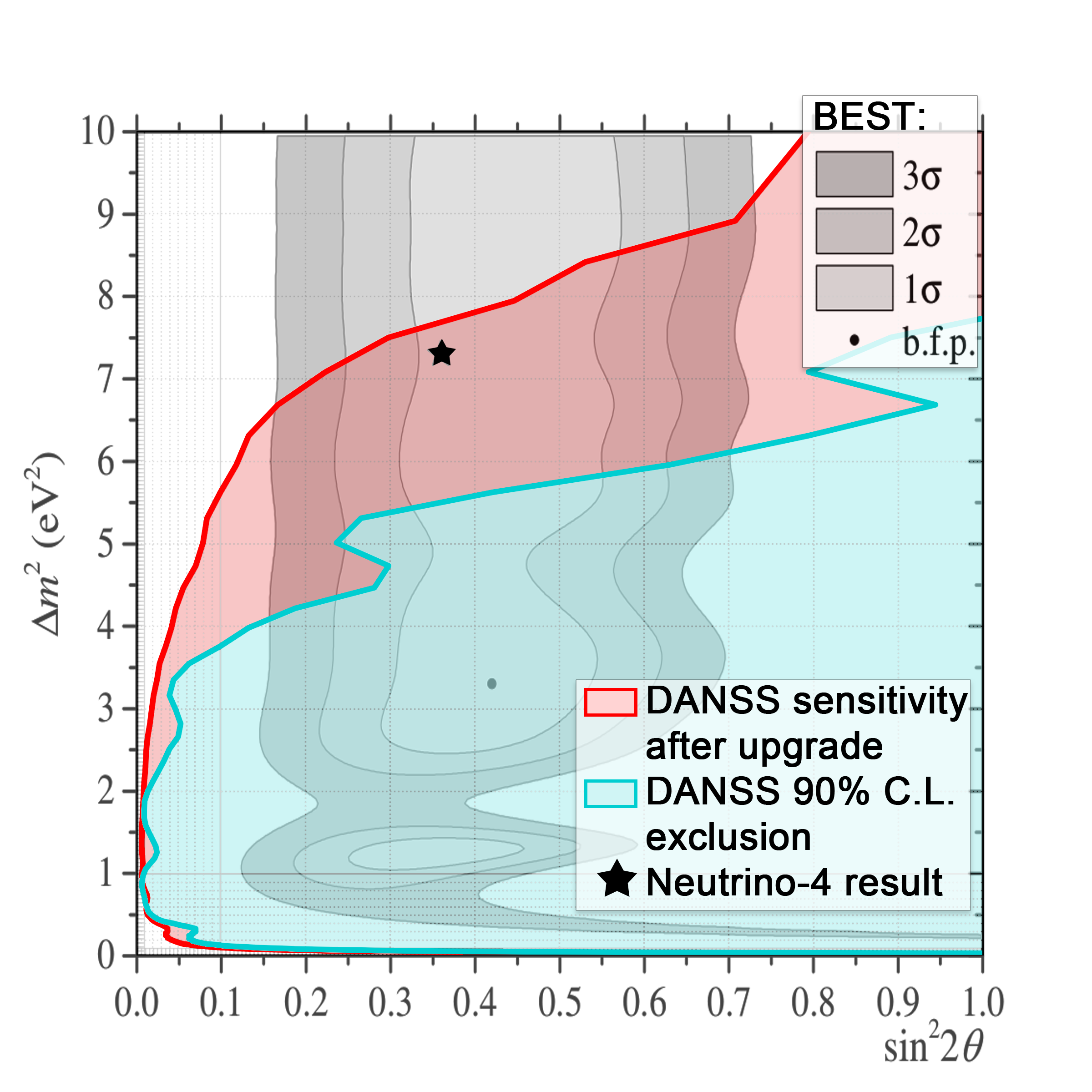}
 \caption{\label{fig:areas} Left panel: DANSS 90\% C.L. exclusion area (cyan) and the boundary of the 90\% C.L. sensitivity area (red dashed). Grey areas show predictions from RAA. Right panel: 90\% C.L. exclusion area for the current analysis (cyan), 90\% C.L. expected sensitivity area after the DANSS upgrade (red). Grey areas show results of the BEST experiment, star marker indicates the best-fit point in the Neutrino-4 experiment.}
% \end{minipage} 
\end{figure}

Figure~\ref{fig:areas} (left) shows the 90\%  Confidence Level (C.L.) area excluded by DANSS in the $\Delta m_{14}^2,~\sin^22\theta_{ee}$ plane. 
The excluded area covers a large fraction of regions indicated by the RAA. In particular, the best fit point $\Delta m_{14}^2=2.3~\rm{eV}^2,~\sin^22\theta_{ee} =0.14$~\cite{Mention2011} is excluded at more than 5$\sigma$ C.L.
A large sterile neutrino parameter region preferred by the recent BEST results~\cite{BEST-2022-arxiv, BEST-2021-arxiv} including the best fit point was already excluded by DANSS and NEOS (see  Figure~\ref{fig:areas}(right)). After modernization in 2022 DANSS will be able to scrutinize much larger area including the Neutrino-4 preferred region (see  Figure~\ref{fig:areas}(right)). Upgraded DANSS will have almost 3 times better energy resolution and 1.7 times larger volume\cite{Sensitivity}.

\subsection{The NEOS Experiment}
The NEOS detector~\cite{Ko:2016owz} had one unsegmented volume  filled with 0.87~ton of 0.5\% Gd-doped liquid scintillator. It was installed at a 2.8~GW$_{th}$ reactor unit 5 of the Hanbit Nuclear Power Complex (Korea) at $23.7\pm 0.3$ m from  the reactor core center. 
As in the DANSS case a quite large  active core size ( 3.1~m in diameter, 3.8~m in height) leads to the oscillation pattern smearing. On the other hand the IBD counting rate is quite high ($\sim$2000 events/day). The background caused by the neutron scattering which imitates the positron signal and subsequent capture of neutrons is rejected with a high efficiency of 73\% using  pulse shape discrimination(PSD) that suppresses signals from the recoil protons. Together with the sizable overburden of $\sim$20~m.w.e.
this allows to achieve a very good S/B ratio of 22.

\begin{figure}[th]%1
\centering
\includegraphics[width=0.95\linewidth]{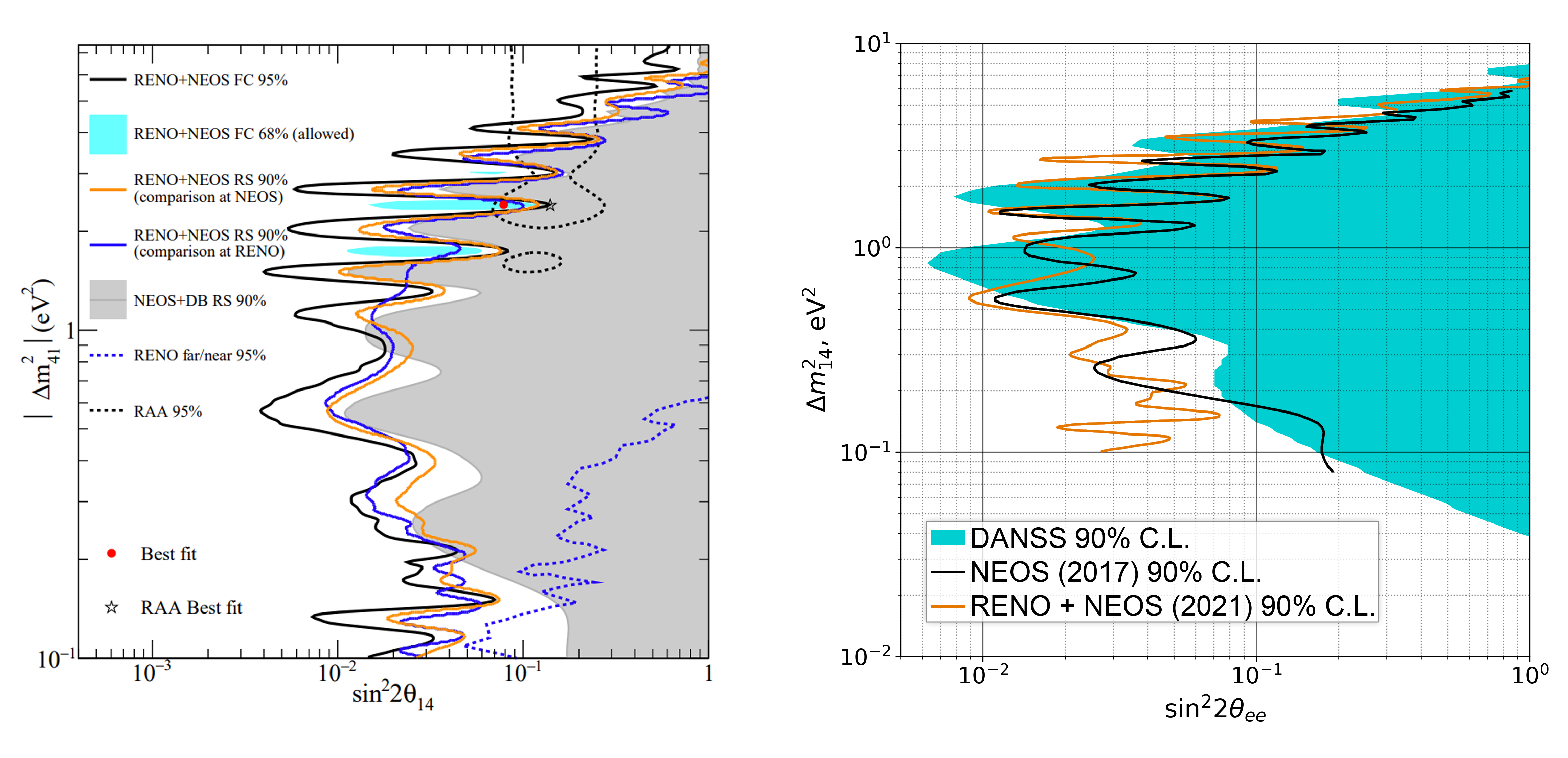}

 \caption{ Left: Comparison of the exclusion limits on sterile neutrino oscillations and an allowed region. The right side of each contour indicates an excluded region. The black curve (cyan filled region) represents a 95\% (68\%) C.L. exclusion contour (allowed region) obtained from the RENO and NEOS combined search using the Feldman and Cousins method~\cite{FeldmanCousins}. The orange (blue) curve represents a 90\% C.L. exclusion contour obtained from the RENO and NEOS combined search using the raster scan method where the spectral comparison is made at NEOS (RENO) detector. The best fit parameter (black point) is found at $|\mydm| = 2.41~\rm{eV}^2$ and $\mysin=0.08$. For the comparison, shown are the NEOS+Daya Bay~\cite{Ko:2016owz} 90\% C.L. (gray shaded) and RENO far/near~\cite{PhysRevLett.125.191801} 95\% C.L. (blue dotted) limits on the disappearance. Also shown is a 95\% C.L. allowed region of RAA~\cite{Mention2011} (black dotted) with the best fit~\cite{WhitePaper} (star). 
 %at $\mysin=0.14$, $\mydm = 2.4~\rm{eV}^2$.
Figure is adopted from~\cite{NEOS2021}.
Right: 90\% C.L. exclusions obtained with the raster scan method. Cyan area -- DANSS exclusions based on 5 millions of IBD events, black line NEOS 2017 results~\cite{Ko:2016owz}, orange line -- combined NEOS + RENO results~\cite{NEOS2021}.}
 \label{NEOSLIM}
\end{figure}

NEOS took data only at one distance from the reactor. Initially NEOS  normalized its data on the $\antiparticle\nu_e$ energy spectrum measured at a different reactor by the Daya Bay collaboration and obtained 
strong limits on the sterile neutrino parameters\cite{Ko:2016owz} shown in Figure~\ref{NEOSLIM} (left).
Later the RENO and NEOS collaborations established even stronger limits using a joint analysis of the data from the same reactor complex~\cite{NEOS2021} (see   Figure~\ref{NEOSLIM} (left)). It is not clear why the Feldman-Cousine statistical analysis method~\cite{FeldmanCousins} leads to much stronger limits in comparison with the 
%CLs\cite{CLs} and 
raster scan or CL$_s$~\cite{CLs} methods. Monte Carlo simulations of generic experiments searching for sterile neutrinos predicted comparable sensitivities for these methods~\cite{CLs}. Figure~\ref{NEOSLIM}~(right) shows that the DANSS and NEOS limits obtained with the same raster scan method 
complement each other.
The RENO-NEOS best-fit point $\Delta m_{14}^2=2.41\pm0.03,~\sin^22\theta_{ee}=0.08\pm0.03$ agrees nicely with the RAA best-fit point. However these parameters are already excluded by DANSS.

In the phase-II NEOS collected data during the whole reactor cycle of 500 days and 2 reactor-off periods. There was some instability in the liquid scintillator response. The data analysis is ongoing.

\subsection{The Neutrino-4 Experiment}

The Neutrino-4 detector is made  of 50 liquid scintillator sections with a total (fiducial) volume of $1.8 (1.42)~{\rm m}^3$~\cite{Neutrino-4-2021}. The detector is installed on a movable platform near a very compact ($42\times42\times35~{\rm cm}^3$) and powerful (100~MW) SM-3 research reactor at Dmitrovgrad (Russia). The distance to the reactor core is changed every 10-15 days which allows to perform measurements in the range from 6~m to 12~m. This is an enormous asset in the control of systematic uncertainties.
A small overburden of 3.5~m.w.e.
%of water equivalent 
and absence of PSD 
lead to a very modest S/B ratio of 0.54. The energy resolution is also modest ($\sim$ 16\% at 1~MeV). In analysis the energy resolution is assumed to be equal to 250~keV for all energies. Such unusual energy independence of the resolution is inferred from the background shape that is dominated by several $\gamma$ lines. However the resolution for $\gamma$ sources that is dominated by measurements of several Compton electrons can be different from the resolution for IBD positrons.

\begin{figure}[th]%1
%\vspace{-4.2cm}
\centering
\includegraphics[width=0.75\linewidth]{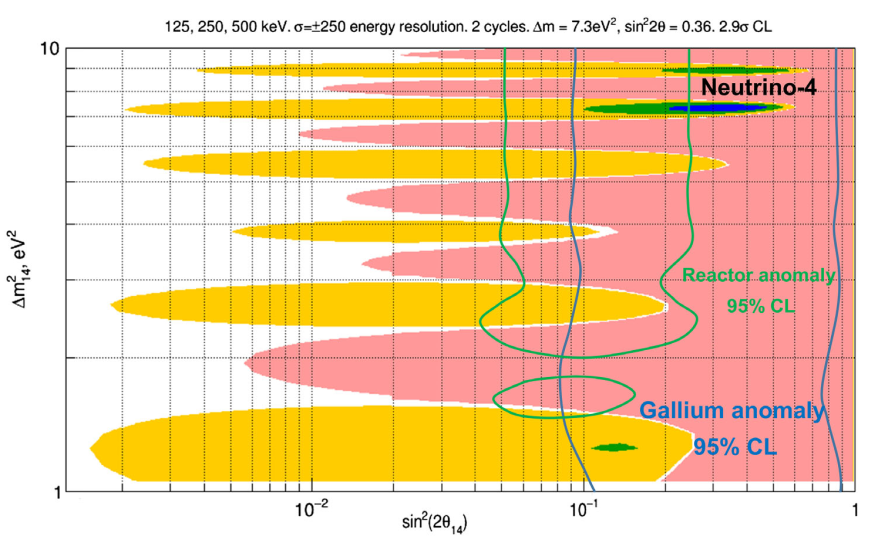}
 \caption{ Accepted (1$\sigma$ -blue, 2$\sigma$ -green, 3$\sigma$ - yellow), and excluded (3$\sigma$ -pink) areas for 3+1 neutrino oscillations obtained by Neutrino-4. Figure is adopted from~\cite{Neutrino-4-2021}.}
 \label{Nu4LIM}
\end{figure}

There were several critical remarks on the Neutrino-4 analysis~\cite{Danilov:2018dme,Danilov:2020rax,Almazan:2020drb,Giunti:2021iti}. Two of them were taken into account in the recent analysis (see Figure~\ref{Nu4LIM})~\cite{Neutrino-4-2021}. The  results were obtained assuming that the test statistics has the $\chi^2$ distribution with 2 degrees of freedom. Since this assumption is only approximately valid Neutrino-4 performed Monte Carlo studies and obtained the 2.7$\sigma$ significance of the sterile neutrino signal including systematic uncertainties. The best fit point is $\Delta m^2_{41} = 7.3\pm 1.17$ eV$^2$, $\sin^2 2\theta_{ee} = 0.36\pm0.12_{\rm stat}$~\cite{Neutrino-4-2021}. 
The obtained sterile neutrino parameters are in tension  with the limits obtained by the reactor $\antiparticle\nu_e$ flux measurements at larger distances even when the large uncertainties in the predictions of the flux are taken into account (see, for example~\cite{MINOS:2020iqj}). 
The PROSPECT and STEREO results~\cite{PhysRevD.103.032001,PhysRevD.102.052002} are also in tension with the Neutrino-4 claim as will be discussed below. The analysis of solar neutrino data excludes $\sin^2 2\theta_{ee} > 0.22$ at 95\% C.L.~\cite{Giunti:2021iti} that is a large fraction of the Neutrino-4 preferred area including the best-fit point.

The Neutrino-4 collaboration is constructing a new detector that will have 3 times better sensitivity~\cite{Neutrino-4-2021}. The new liquid scintillator with larger amount of Gd will
decrease the neutron capture time and hence the background. PSD will further decrease the background. The energy resolution will be considerably improved by using a two-sided readout of horizontal sections (now they are vertical with one Photo-Electron Multiplier (PMT)). The new detector should be completed already in 2022.

\subsection{The PROSPECT experiment}
\begin{figure}[th]%1
%\vspace{-4.2cm}
\centering
\includegraphics[width=0.49\linewidth]{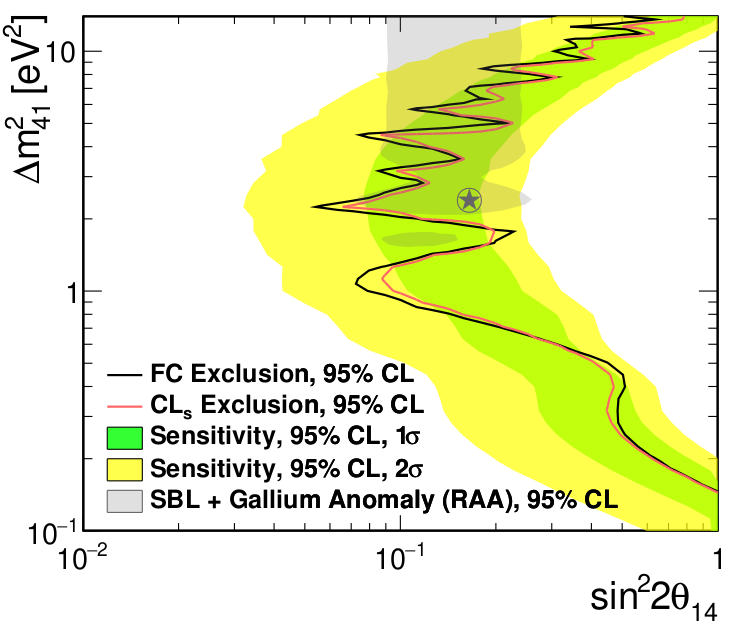}
\includegraphics[width=0.44\linewidth]{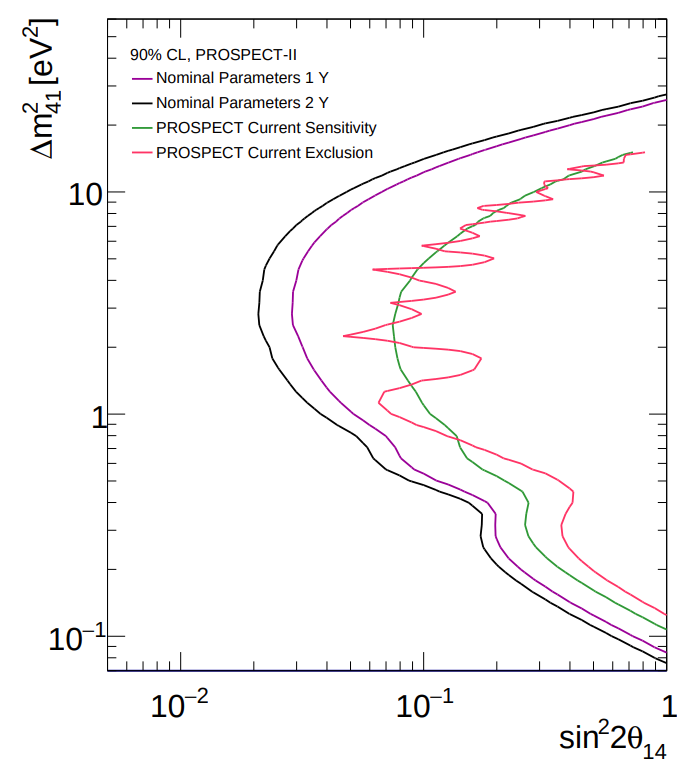}

 \caption{
Left: Oscillation exclusion contours derived using the Gaussian CL$_s$ and Feldman-Cousins (FC) methods. Also pictured are the 1$\sigma$ and 2$\sigma$ (green and yellow) exclusion ranges produced by PROSPECT toy MC data sets, as well as the RAA preferred parameter space and best-fit point from Ref~\cite{Mention2011}.
Figure is adopted from~\cite{PhysRevD.103.032001}. %(figure 45).}
Right: Comparison of sterile oscillation sensitivities for different current and
projected PROSPECT~\cite{PhysRevD.103.032001} and PROSPECT-II data sets. Figure is adopted from~\cite{arXiv:2107.03934}.
}
\label{PROSPECT_LIM}

\end{figure}

The PROSPECT detector~\cite{PROSPECT_2018, PhysRevD.103.032001}  consists of
154 optically isolated rectangular segments (14.5~cm$\times$14.5~cm$\times$117.6~cm) filled with 
liquid scintillator  loaded with $^6$Li in order to capture and detect neutrons and read out by two PMT each. The PROSPECT detector is installed at the High Flux Isotope Reactor (HFIR) at Oak Ridge National Laboratory with less than 1~m.w.e. 
%one meter-water-equivalent of 
overburden. Nevertheless, a very good PSD and 3D reconstruction of events allowed PROSPECT to achieve a decent S/B ratio of 1.36. A high reactor power (85~MW), the large detector ($\sim4$~ton), and a small distance from the reactor (6.7~m) allow PROSPECT to collect about 800 IBD events per day with excellent energy resolution of 4.5\% at 1~MeV.
Unfortunately 42\% of segments were not operational because of problems with PMTs.
Nevertheless PROSPECT  excluded a sizable part of the sterile neutrino parameters (see Figure~\ref{PROSPECT_LIM})~\cite{PhysRevD.103.032001}. The obtained limits are in tension with the Neutrino-4 result but can not exclude it. 

PROSPECT plans to upgrade the detector and increase considerably the sensitivity (see Figure~\ref{PROSPECT_LIM} right)~\cite{arXiv:2107.03934}.

\subsection{The SoLid experiment}

The extremely highly segmented SoLid detector consists of 12800 polyvinyltoluene cubes ((5$\times$5$\times$5)~cm$^3$)
with thin sheets of $^6$LiF:ZnS(Ag) to capture and detect neutrons~\cite{SoLid2020}. Light from each cube is collected with three orthogonal wavelenght-shifting fibers read out with Silicon Photo Multipliers (SiPM). The SoLid detector is installed at a distance of $\sim$~6~m of the 60~MW SCK$\cdot$CEN BR2 research reactor in Belgium. It has a modest energy resolution of $\sim$14\% at 1~MeV and good PSD for background rejection. Unfortunately the internal radioactivity from the contamination of $^6$LiF:ZnS layers leads to a very high background. With complicated machine learning techniques SoLid managed to extract the IBD signal but so far has no results on sterile neutrino searches. SoLid upgrades the detector with new SiPMs to mitigate the problems~\cite{SolidConf}. 

\subsection{The STEREO experiment}
\begin{figure}[th]%1
%\vspace{-4.2cm}
\centering
\includegraphics[width=0.45\linewidth]{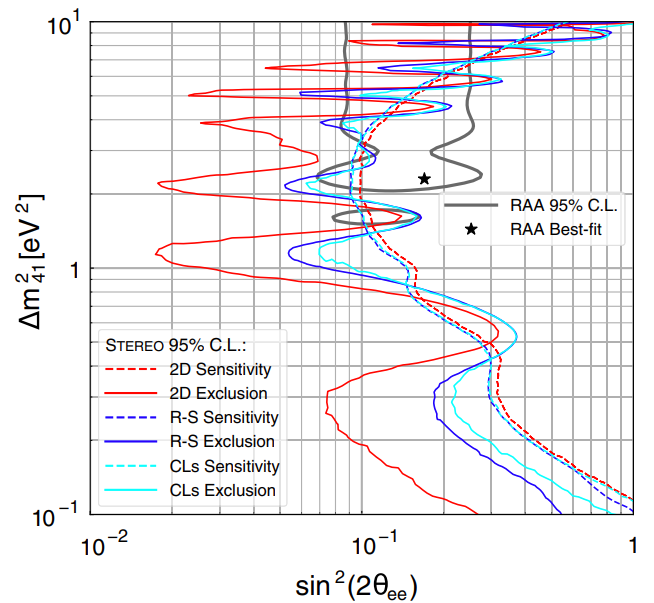}
 \caption{\footnotesize 
Comparison of the exclusion contours (solid) and exclusion sensitivity contours (dashed) at 95\% C.L. of phase I + II, for the two-dimensional method (red), the raster-scan method (dark blue), and the CL$_s$ method (light blue). Overlaid are the allowed regions of the RAA (grey) and its best-fit
point (star) \cite{WhitePaper}. Figure is adopted from~\cite{PhysRevD.102.052002}.}
 \label{STEREO}
\end{figure}

The STEREO detector~\cite{PhysRevD.102.052002} is made of six optically separated cells
%of the target volume 
filled with a gadolinium loaded liquid scintillator.
STEREO was installed at the High Flux Reactor of the Institute Laue-Langevin. The cell distances from the core range from 9.4~m to 11.1~m.
Elaborated PSD technique was used to suppress the background from fast neutrons. Still, the achieved S/B ratio of 0.9 was quite modest.
STEREO excluded a sizable fraction of the sterile neutrino parameter space~\cite{PhysRevD.102.052002} (see Figure~\ref{STEREO}). As in the case of NEOS, the Feldman-Cousines analysis method gives much stronger limits in comparison with other methods. Moreover, the limits are much stronger than the experiment sensitivity.

STEREO finished the data taking but the data analysis is still ongoing.
 
\subsection{The JUNO-TAO experiment}
The Juno-TAO is a ton-level
liquid scintillator detector at $\sim$ 30 meters from a reactor core of the  Taishan Nuclear Power Plant in Guangdong, China~\cite{JUNO:2020ijm}. The detector will have extremely good energy resolution of 2\% at 1~MeV and will detect 2000~$\antiparticle\nu_e$ per day. The expected Juno-TAO 90\% C.L. sensitivity on $\sin^22\theta_{ee}$ with 3 years of data taking is about 0.01-0.02 in the wide range of 
$\Delta m_{14}^2$  between  0.06~eV$^2$ and 3~eV$^2$.
%(see Figure\ref{I-7 TAO}??
JUNO-TAO plans to start data taking in 2022. 
\subsection{Summary on Very Short Base-Line (VLBS) experiments}
 The advantages (red color) and disadvantages (blue color) of recent VSBL experiments are summarized in Table~\ref{Table}.
The experiments at industrial reactors (DANSS and NEOS) benefit from the high counting rates up to 5000~$\antiparticle\nu_e$/day. They have the highest sensitivity of $\sin^22\theta_{ee} < 0.01$ at $\Delta m_{14}^2\approx$~(1-2)~eV$^2$. 
At larger $\Delta m_{14}^2$ oscillations are averaged out already inside the large reactor core and the sensitivity deteriorates. For large $\Delta m_{14}^2$ the experiments at the research reactors in particular upgraded Neutrino-4 and PROSPECT have higher sensitivity. 
The upgraded DANSS, Neutrino-4, and PROSPECT experiments will soon test the Neutrino-4 claim of the sterile neutrino observation and scrutinize even larger part of the sterile neutrino parameters preferred by the recent BEST results.

\begin{table}[h]
\caption{\label{Table}Parameters of the VSBL experiments. Advantages and disadvantages are indicated with the red and blue colors correspondingly.} 
\begin{center}
\lineup
\begin{tabular}{*{7}{c}}
\br                              
 & DANSS & NEOS &Neutrino-4 & PROSPECT & SoLid & STEREO\\
\mr
Power [MW] & \textcolor{goodcolor}{3100} & \textcolor{goodcolor}{2815} & \textcolor{goodcolor}{100} & \textcolor{goodcolor}{85} & 50-80 & 58\\
Core size [cm] & \textcolor{badcolor}{$\diameter = 320$}  & \textcolor{badcolor}{$\diameter = 310$}  & \textcolor{goodcolor}{$42\times42$} & \textcolor{goodcolor}{$\diameter = 51$} & \textcolor{goodcolor}{$\diameter = 50$} & \textcolor{goodcolor}{$\diameter = 40$}\\
 &\textcolor{badcolor}{$h = 370$}  & \textcolor{badcolor}{$h = 380$} & \textcolor{goodcolor}{$h = 35$} & \textcolor{goodcolor}{$h = 44$} & \textcolor{goodcolor}{$h = 90$} & \textcolor{goodcolor}{$h = 80 $}\\
Overburden [mwe] & \textcolor{goodcolor}{50} & \textcolor{goodcolor}{20} & 3.5 & $<1$   & 10  & 15  \\
Distance [m] & 10.9-12.9 & \textcolor{badcolor}{23.7} & \textcolor{goodcolor}{6-12} & \textcolor{goodcolor}{7-9} & \textcolor{goodcolor}{6-9} & 9-11\\
 &  \textcolor{goodcolor}{movable} &  & \textcolor{goodcolor}{movable} & & & \\
IBD events/day & \textcolor{goodcolor}{5000} & \textcolor{goodcolor}{2000} & \textcolor{badcolor}{200} & 750 & ? & 400\\
PSD & \textcolor{badcolor}{No} & \textcolor{goodcolor}{Yes} & \textcolor{badcolor}{No} & \textcolor{goodcolor}{Yes} & \textcolor{goodcolor}{Yes} & \textcolor{goodcolor}{Yes}\\
Readout & \textcolor{goodcolor}{Quasi-3D} & \textcolor{badcolor}{1D} & 2D & \textcolor{goodcolor}{3D} & \textcolor{goodcolor}{3D} & 2D\\
S/B & \textcolor{goodcolor}{50} & \textcolor{goodcolor}{23} & \textcolor{badcolor}{0.54} & 1.36 & ? & \textcolor{badcolor}{0.9}\\
$\sigma_E/E$~[\%] at 1 MeV & \textcolor{badcolor}{34} & \textcolor{goodcolor}{5} & \textcolor{badcolor}{16} & \textcolor{goodcolor}{4.5} & \textcolor{badcolor}{14} & 8\\
\br
\end{tabular}
\end{center}
\end{table}

\section{Global fits}

\begin{figure}[th]%1
%\vspace{-4.2cm}
\centering
\includegraphics[width=0.45\linewidth]{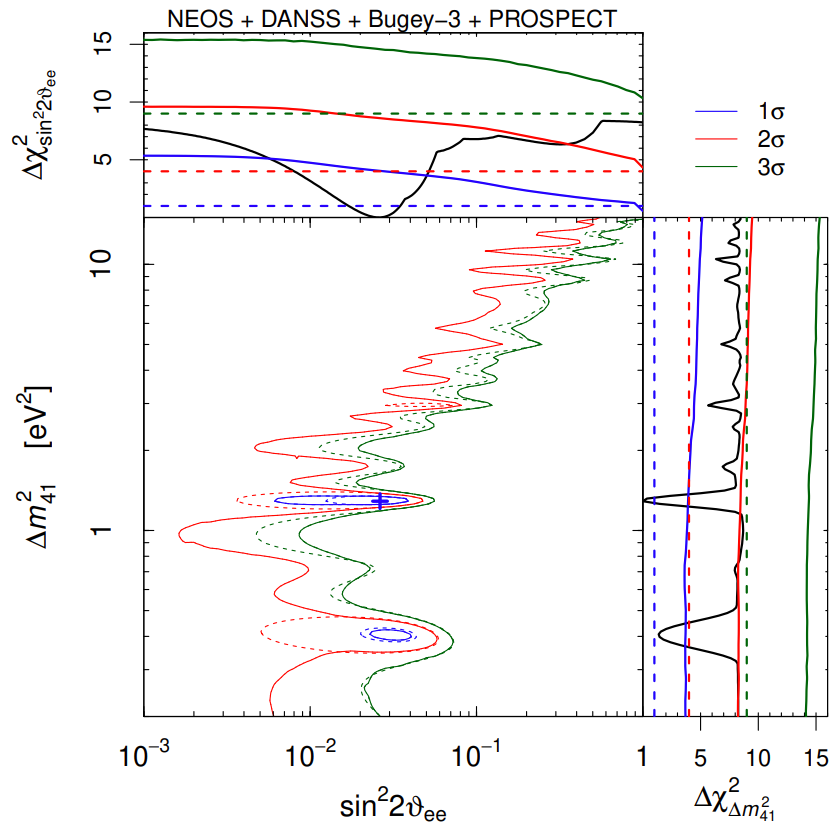}
 \caption{
 Contours of the 1$\sigma$ (blue), 2$\sigma$ (red), and 3$\sigma$ (green) allowed regions in the (\oscillationspars) plane obtained with the combined analysis~\cite{PhysRevD.101.095025} of the data of the four reactor spectral ratio experiments NEOS~\cite{Ko:2016owz}, DANSS~\cite{arXiv:1911.10140}, Bugey-3~\cite{Bugey_2018}, and PROSPECT~\cite{PROSPECT_2018}. The solid lines represent the contours obtained with Monte Carlo evaluation of the distribution of $\Delta\chi^2$, and the dashed lines depict the contours obtained
assuming the $\chi^2$ distribution. Also shown are the marginal
$\Delta\chi^2$’s (black) for \oscillationspars\ together with the $\Delta\chi^2$
values corresponding to 1$\sigma$ (blue), 2$\sigma$ (red), and 3$\sigma$ (green)
obtained with the $\chi^2$  (dashed) and Monte-Carlo (solid) distributions. The blue cross indicates the best-fit point. Figure is adopted from~\cite{PhysRevD.101.095025}. 
}
 \label{fig:GlobalGiunti}
\end{figure}

It is not trivial to combine the results of different experiments on sterile neutrino searches. Initially this was done (see e.g.~\cite{Dentler_2018, Giunti_2018}) assuming that the test statistics $\Delta\chi^2= \chi^2(3\nu)-\chi^2(4\nu)$ follows the $\chi^2$ distribution with 2 degrees of freedom which is not corrects in the sterile neutrino searches (see e.g.~\cite{PhysRevD.101.095025}). The Monte Carlo simulation of the $\Delta\chi^2$ distribution is needed for correct estimates of the significance that requires deep knowledge of experiments. To circumvent this very serious complication several authors performed  a simplified simulation of experiments used in their analysis~\cite{PhysRevD.101.095025,arXiv:2005.01756v2}. With the Monte Carlo determination of the $\Delta\chi^2$ distribution the significance of the evidence for sterile neutrinos in $\nu_e/\antiparticle\nu_e$ disappearance experiments (without Neutrino-4 and BEST) was reduced from 2.4$\sigma$ to 1.8$\sigma$ (see Figure~\ref{fig:GlobalGiunti})~\cite{PhysRevD.101.095025}.

\begin{figure}[th]%1
%\vspace{-4.2cm}
\centering
\includegraphics[width=0.45\linewidth]{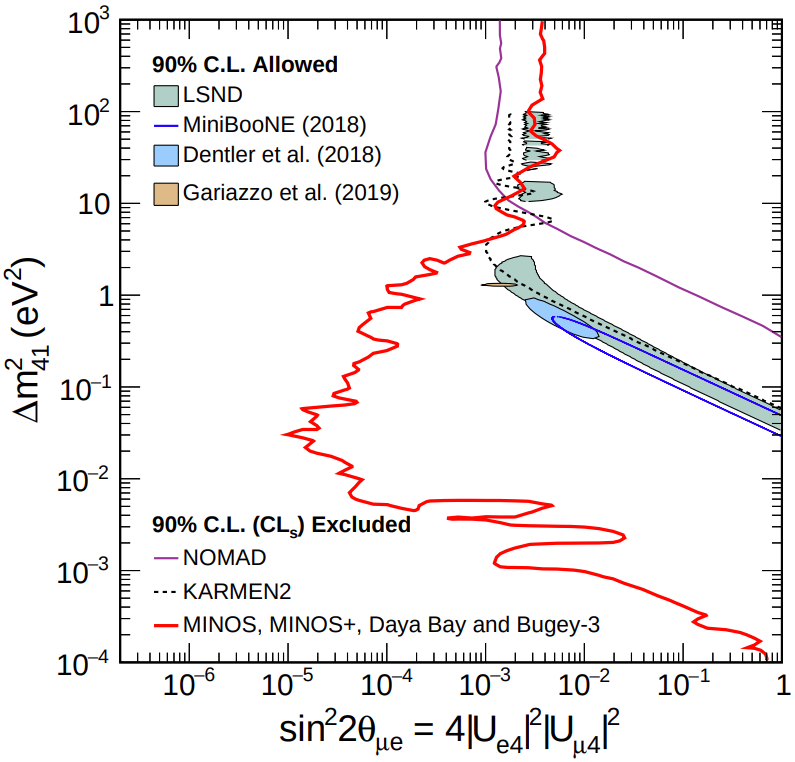}
 \caption{
 Comparison of the MINOS, MINOS+, Daya Bay, and Bugey-3 combined 90\% CLs limit on $\sin^2 2\theta_{e\mu}$ to the LSND and MiniBooNE 90\% C.L. allowed regions. Regions of parameter space to the right of the red contour are excluded. The regions excluded at 90\% C.L. by the KARMEN2 Collaboration~\cite{PhysRevD.65.112001} and the NOMAD Collaboration~\cite{ASTIER200319} are also shown. The combined limit also excludes the 90\% C.L. region allowed by a fit to global data by Gariazzo et al. where MINOS, MINOS+, Daya Bay, and Bugey-3 are not included~\cite{Giunti_2018, GARIAZZO201813}, and the 90\% C.L. region allowed by a fit to all available appearance data by Dentler et al.~\cite{Dentler_2018} updated with the 2018 MiniBooNE appearance results~\cite{MiniBooNE2_2018}.
Figure is adopted from~\cite{MINOS:2020iqj}.}
 \label{fig:MinosDBBugey}
\end{figure}

The strong limits on $\antiparticle\nu_e$ and $\nu_\mu$ disappearance  do not allow to   explain by sterile neutrinos simultaneously the LSND/MiniBooNE excess in the ($\antiparticle\nu_e$)$\nu_e$ appearance and RAA/GA deficit(see Figure~\ref{fig:MinosDBBugey})~\cite{MINOS:2020iqj}. However, the Neutrino-4 result and the highly significant BEST result were not included into the comparison.

\begin{figure}[th]%1
%\vspace{-4.2cm}
\centering
\includegraphics[width=0.45\linewidth]{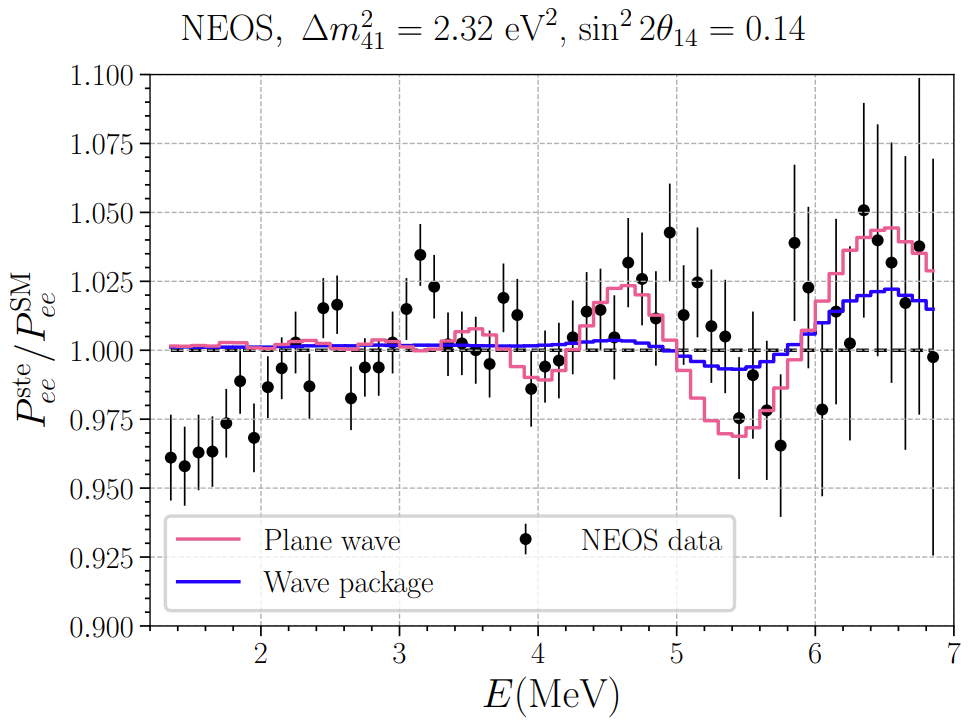}
 \caption{
 %Example of the effect in NEOS. 
 Illustration  of the decoherence effect, for $\sigma_x = 2.1 \times 10^{-4}$~nm(blue curve), with the reactor antineutrino anomaly (RAA) best-fit parameters:  $\Delta m_{14}^2=2.32~\rm{eV}^2,~\sin^22\theta_{ee} =0.14$ (red curve). In the y-axis, the ratio between
 expected number of events in 4$\nu$ and 3$\nu$ models as well as the experimental results from
 NEOS~\cite{Ko:2016owz}.
 Figure is adopted from~\cite{Packet}.
}
 \label{fig:packet}
\end{figure}

The validity of existing limits on sterile neutrino parameters was questioned recently~\cite{Packet}. For very small neutrino wave packet sizes evolution of the neutrino beams with time becomes more complicated than equation (\ref{eq:1-osc_ee}) and the oscillation behavior is smeared out (see Figure~\ref{fig:packet}). Existing experimental limits on the size of  reactor antineutrino wave packets  are quite weak. If one uses the present upper limits as the actual values of the reactor antineutrino wave packet size, the limits on the sterile neutrino mixing parameter $\sin^22\theta_{ee}$ %($\mysin$) 
become considerably weaker and the tension between the results for neutrino disappearance  in radioactive sources experiments and reactor experiments vanishes at least for some values of $\mydm$.
The contradiction between appearance and disappearance experiments is also reduced. However there is no reason for the wave packet size to be equal to its current upper limit. Most probably the size is much larger and it does not change the present limits on the sterile neutrino parameters.

The standard cosmological model strongly disfavors the eV-scale sterile neutrinos. However there are models which include eV-scale  sterile neutrinos and reduce the tension between different determinations of the Hubble constant~\cite{Archidiacono_2020}.

\section{Conclusions}

\begin{figure}[th]%1
%\vspace{-4.2cm}
\centering
\includegraphics[width=0.45\linewidth]{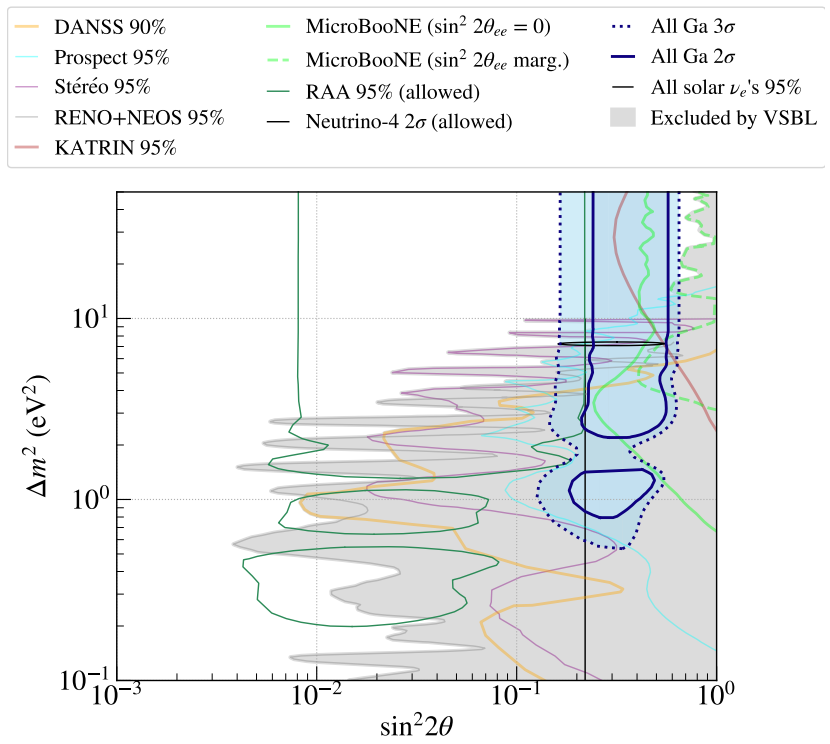}
 \caption{Exclusion contours of all gallium anomaly experiments: two GALLEX, two SAGE and two BEST results. The blue solid line and the blue dotted line show the 2$\sigma$ and 3$\sigma$
confidence level, respectively. The figure also presents the exclusion contours from Prospect~\cite{PhysRevD.103.032001}, DANSS~\cite{SkrobovaICNFP2019}, Stereo~\cite{PhysRevD.102.052002}, KATRIN~\cite{PhysRevLett.126.091803}, the combined analysis of RENO and NEOS data~\cite{NEOS2021}, reactor anti-neutrino anomalies (RAA)~\cite{Mention2011}, interpretations of the MicroBooNE result for the oscillation hypothesis with fixed mixing angle ($\sin^2 2\theta$) and profiled over the angle~\cite{2111.10359v1}, and the model-independent 95\% upper bound on $\sin^2 2\theta$ from all solar neutrino experiments~\cite{Giunti:2021iti}. The 2$\sigma$ allowed region of Neutrino-4~\cite{Serebrov:2020rhy} is also presented and the grey shading represents the merged exclusion of the very short baseline (VSBL) null results. Figure is adopted from~\cite{BEST-2022-arxiv}.
}
 \label{fig:BestEclusions}
\end{figure}

The situation with the experimental indications of eV-scale sterile neutrino existence is quite controversial. 
Indications from all but one reactor experiments have statistical significance below 2$\sigma$. This should be compared with the initial estimates of the RAA significance of about 3$\sigma$. On the other hand Neutrino-4 claims the observation of sterile neutrinos with 2.7$\sigma$ significance. Moreover the large $\nu_e$ deficit observed recently by BEST with  more than 5$\sigma$ significance perfectly agrees with the Neutrino-4 result. However a large fraction of sterile neutrino parameters preferred by Neutrino-4 and BEST are either excluded or are in tension with results from several experiments. This is nicely illustrated by Figure~\ref{fig:BestEclusions}~\cite{BEST-2022-arxiv}.

Upgraded DANSS, Neutrino-4, and PROSPECT experiments as well as JUNO-TAO will clarify the situation in a few years.

More than 6$\sigma$ evidence  for $\nu_e$($\antiparticle\nu_e$) appearance from LSND/MiniBooNE is not confirmed by MicroBooNE, but not excluded completely.
The Fermilab Short Baseline Neutrino program 
and the JSNS$^2$ experiment at J-PARK~\cite{AJIMURA2021165742}
will clarify the situation in a few years.

The next several years will be crucial for the searches for eV-scale sterile neutrinos.

\section{Acknowledgments}
Author would like to thank N.Skrobova for the invaluable help in the preparation of the manuscript. 
This work is supported by the Ministry of Science and Higher Education of the Russian Federation under the Contract No. 075-15-2020-778.

\end{document}